# Information Leaks via Safari's *Intelligent Tracking Prevention*

ARTUR JANC, KRZYSZTOF KOTOWICZ, LUKAS WEICHSELBAUM, ROBERTO CLAPIS

## ABSTRACT

*Intelligent Tracking Prevention (ITP) is a privacy mechanism implemented by Apple's Safari browser, released in October 2017[1]. ITP aims to reduce the cross-site tracking of web users by limiting the capabilities of cookies and other website data[2].*

*As part of a routine security review, the Information Security Engineering team at Google has identified multiple security and privacy issues in Safari's ITP design. These issues have a number of unexpected consequences, including the disclosure of the user's web browsing habits, allowing persistent cross-site tracking, and enabling cross-site information leaks[3] (including cross-site search[4]).*

*This report is a modestly expanded version of our original vulnerability submission to Apple (WebKit bug #201319[5]), providing additional context and edited for clarity. A number of the issues discussed here have been addressed in Safari 13.0.4 and iOS 13.3, released in December 2019[6].*

## 1 BACKGROUND

### 1.1 Intelligent Tracking Prevention (ITP)

The aim of Safari's Intelligent Tracking Prevention mechanism is to protect users from tracking across the web by preventing websites commonly loaded in a third-party context from receiving information which would allow them to identify the user. This functionality has two core elements:

- Establishing an on-device list of *prevalent domains* based on the user's web traffic.
- Applying privacy restrictions to cross-site requests to domains designated as prevalent.

When Safari notices a website sending a cross-site resource request, it increases an internal counter for the domain from which the resource is loaded (referred to as an *ITP strike* throughout this report). Once a given domain has accumulated enough ITP strikes, it is categorized by Safari as a prevalent domain. Details of the classification logic evolve and are beyond the scope of this document; in our testing, being used in a third-party context by **3** other domains was consistently sufficient for Safari to designate a domain as prevalent.

The prevalent domain list (referred to as the *ITP list* below) is stored at the granularity of registrable domains; specifically, Safari stores the eTLD+1, accounting for the Public Suffix List[7].

When Safari makes a cross-site request to a prevalent domain, it applies privacy restrictions to remove information that would allow that domain to infer the user's identity and cross-link it with third-party requests from other websites. It does so by removing cookies and truncating the `Referer` header to include only the referring document's origin instead of its full URL.

The ITP list is append-only, but it is cleared whenever the user clears their Safari browsing history; the entire list is wiped even if the user resets history for a short time period. Private Browsing Mode does not reuse the ITP list from the main browsing profile.

### 1.2 Attack overview

As a result of customizing the ITP list based on each user's individual browsing patterns, Safari has introduced global state into the browser, which can be modified and detected by every document.

Any site can issue cross-site requests, increasing the number of ITP strikes for an arbitrary domain and forcing it to be added to the user's ITP list. By checking for the side effects of ITP triggering for a given cross-site HTTP request, a website can determine whether its domain is present on the user's ITP list; it can repeat this process and reveal ITP state for any domain.

Because the ITP list implicitly stores information about the websites visited by the user, leaking its state reveals sensitive private information about the user's browsing habits. Attacks #1 and #2 focus on this vulnerability. Attack #3 demonstrates how to use the ITP list to create a persistent fingerprint that will follow the user around the web. Attacks #4 and #5 demonstrate the potential for malicious exploitation of ITP to affect or reveal cross-origin application state.

### 1.2.1 Checking if a domain is on the ITP list.

A key component of several of the described attacks is the ability to determine if an arbitrary domain is on the user's ITP list; we discuss this before moving on to the details of the attacks.

It is of course trivial to detect the ITP status of any domains under the attacker's control: the attacker can directly issue cross-site requests from another domain and inspect them for the effects of applying ITP restrictions. For example, the attacker can check if the `Referer` header has been truncated, or if a cookie previously set in a first-party context was present in the request; ITP's design can do little to prevent this capability.

```
<!-- Document with a long URL: https://attacker.example/<16000 bytes>/attack -->

<script>
document.querySelectorAll('img').forEach(img => {
  img.onload = function() { /* Successful load indicates ITP triggered */ }
  img.onerror = function() { /* Unsuccessful load indicates no ITP */ }
})
</script>
```

Listing 1. Setting up resource loads for ITP detection

To reveal the ITP status of domains *outside* of their control, the attacker must use a side-channel to distinguish the behavior of requests affected by ITP from those that were not subject to ITP restrictions. Importantly, the web abounds in such side channels. For example, an attacker can:

- Send a request with an overly long `Referer` header, which will be rejected by the server if the full `Referer` is present, but allowed when it's truncated. Most webservers enforce a maximum GET request size limit between 8KB and 128KB, and reject larger requests. An attacker can select any resource fetchable in `no-cors` mode (e.g. image, script or stylesheet) and use the `onload`/`onerror` events to learn whether the load succeeded, revealing whether the hosting domain was on the user's ITP list; Safari currently has no `Referer` size limit. This is the technique we chose for our proofs-of-concept; it was successful on all 100+ domains in our testing.
- Request a `no-cors` resource which requires authentication. If this results in an error as a result of missing credentials, the attacker can infer the domain's presence on the ITP list.
- Send a request to an open redirector which will issue a 30x redirect to the attacker's site. The attacker can use Content Security Policy to detect the cross-origin redirect, or check for cookies in the incoming request. This functionality is common in tracking domains.
- Send a request which will issue a redirect in the absence of credentials in the request. The redirect can be detected using Fetch's `redirect:'manual'`.
- Identify endpoints which host user-controlled files and upload a document which reads `document.referrer`. This functionality is common in advertising domains.
- A local network attacker or ISP can also directly observe whether cookies or the `Referer` header were present in a plaintext HTTP request, without relying on any other techniques.

```
1 GET /favicon.ico HTTP/1.1
2 Host: itp.example
3 Referer: https://attacker.example
4 ...
5
6 HTTP/1.1 200 OK
7 Content-Type: image/png
8 ...
```

Listing 2. Example successful request to an ITP domain

```
1 GET /favicon.ico HTTP/1.1
2 Host: non-itp.example
3 Referer: https://attacker.example/<16000 bytes>/attack
4 Cookie: NON_ITP_COOKIE=value;
5 ...
6
7 HTTP/1.1 413 Request entity too large
8 Content-Type: text/html
9 ...
```

Listing 3. Example failed request to a non-ITP domain

This list of side channels above is not exhaustive and it's very likely that additional techniques to reveal the ITP status of arbitrary domains exist. Importantly, given that ITP is tracked at the eTLD+1 level, the attacker can pick an arbitrary URL under the tested domain that meets the attacker's conditions. In our testing, every domain allowed one or more techniques to check for its ITP status. This enabled the general attacks mentioned in the next section.

## 2 ATTACKS

### 2.1 Attack #1: Revealing domains on the ITP list

PoC #1: https://itptesting.appspot.com/static/01-reveal-itp-domains.html

Using the side channels discussed above an attacker can learn the ITP status of an arbitrary domain (PoC #1 uses overlong `Referer` headers). This indirectly reveals information about the user's browsing habits: many domains are used for cross-site fetches from a particular set of sites (e.g. agora.pl is on the ITP list for viewers of news sites in Poland, and phncdn.com for visitors of certain adult sites), or based on other criteria (e.g. google.ch is in the ITP list for users located in Switzerland).

Given that the ITP list contains implicit information about the user's entire browsing history since they last cleared their browsing data, the state of the list can reveal surprisingly sensitive information about the user. For example, in a hypothetical scenario where there are only 3 websites that make cross-site requests to a given domain, the presence of that particular domain on the ITP list discloses that the user had visited all the requesting sites. Attack #2 refines this approach to infer even more information about the user's browsing state.

The ITP list can additionally be used as a fingerprint of the user. Unrelated sites read use the contents of the ITP list to establish that their visitor is the same person who visited another website, even without an explicit relationship between the sites. This information can also be obtained by limited contexts such as sandbox iframes; it could be used – for example – by an advertiser who wants to establish a link between their ad impression and a user's later visit to the advertiser's site.

## 2.2 Attack #2: Identifying individual visited websites

PoC #2: `https://itptesting.appspot.com/static/02-identify-visited-sites.html`

Many web applications use multiple domains to host resources needed for the operation of the site; common scenarios include hosting user-uploaded files in a separate sandbox domain (e.g. mail-attachment.googleusercontent.com) or using a custom CDN for static resources (e.g. chasecdn.com). This puts an ITP strike on a given domain.

For any domain, the attacker can learn its number of ITP strikes (i.e. how many eTLD+1s sent cross-site requests to that domain) by checking how many additional attacker-controlled eTLD+1s need to request a resource from the domain for it to appear on the ITP list. For example, assuming an ITP threshold of 3 strikes, if a domain needed to be used in a cross-site context 2 times before it appeared on the ITP list, the attacker knows that previously 1 other site has loaded resources from that domain. This increases the granularity of the data inferred about the user's browsing habits and has consequences even for sites which aren't classified by Safari as prevalent domains.

Specifically, it allows an attacker to detect that a user previously visited any site which makes cross-site requests to a custom domain used only by that site.

## 2.3 Attack #3: Creating a persistent fingerprint via ITP pinning

PoC #3: `https://itptesting.appspot.com/static/03-create-itp-fingerprint.html`

An attacker can add their own domain to the user's ITP list by making cross-site requests to it from at least 3 other domains. As discussed in Section 1.2.1, the domain's ITP status can easily be determined on the server by checking for the full `Referer` or `Cookie` headers in incoming cross-site requests.

This allows an attacker to track the user by creating a small number of *pinning domains*, creating a random identifier (e.g. a 32-bit number) and encoding it as a binary value in the (sub)set of the pinned domains. This technique is equivalent to HSTS pinning[8]. As a result, attackers can create a global shared identifier that can be accessed or set from every website, including from limited contexts such as iframes and documents in the HTML sandbox, which otherwise don't have access to persistent storage.

This technique can be further improved to create a more convenient large-scale tracking mechanism. Attackers can use domains on the Public Suffix List to avoid registering separate pin domains. To prevent resetting the fingerprint after reading ITP pins in a cross-site context from a large number of sites (which would put all pin domains on the ITP list), an attacker can recreate such a "saturated" fingerprint on a different set of domains, or conduct the detection via short-lived documents, which do not increase the number of ITP strikes for their cross-site resources.

## 2.4 Attack #4: Forcing a domain onto the ITP list

PoC #4: `https://itptesting.appspot.com/static/04-add-itp-domain.html`

Similarly to Attack #3, it's possible to add an arbitrary domain to the user's ITP list. This can lead to vulnerabilities on domains that never expected to have their requests modified by ITP. For example, if an SSO-enabled application consists of two domains which require login credentials and a document on one domain loads authenticated resources from the other, an attacker can force these requests to fail by making ITP unexpectedly remove cookies.

If the blocked resource provides any sensitive functionality, such as performing security checks, the capability to disable these checks may have security consequences. In other cases, it may result in denial of service until the user resets their browsing history.

This technique can also be used as a stealthier variant of Attack #3 where an attacker adds domains outside of their control to the ITP list, and later uses their presence on the ITP list as a fingerprint.

## 2.5 Attack #5: Cross-site search attacks using ITP

In web applications with search functionality, it's frequently possible for an attacker to open a new window/frame to a search page with a chosen query. If the attacker can distinguish between a response which returned results and an empty one, they can learn information about private search results for a given user; this can lead to damaging attacks, such as exfiltrating partial contents of a user's webmail inbox[4].

ITP introduces an information leak if the search results page includes any cross-site resource when results are present, but omits that resource when no results were found (or vice versa). This is commonly the case in applications which let users search photos or videos, where the media resources are served from a separate domain for security reasons.

By creating 2 artificial ITP strikes for the cross-site domain and checking if a search query resulted in the domain being put on the ITP list, the attacker can leak user state from the application.

## 3 MITIGATIONS

### 3.1 General mitigations

A fundamental issue responsible for the browsing history disclosure attacks discussed in this report is the user-dependent of ITP: customizing the prevalent domain list based on individual browsing patterns. A potential avenue for addressing this problem may be to remove user-specific aspect of ITP by shipping a common list, identical for all Safari users, based on an enumeration of known trackers or federated learning; however, this may make ITP less effective. Ideally, the privacy-protecting behaviors applied to sites on the ITP list could be extended to all domains, applying ITP by default to the entire web; this, in turn, may be prohibitive for compatibility reasons.

### 3.2 Short-term workarounds

Several of the simple, general techniques to detect the state of ITP for non-cooperating domains could be hindered by making small implementation changes, including:

- Limiting the length of the `Referer` header[9] to prevent overlong referring URLs from forcing a server error detectable in a cross-site context.
- Disabling `redirect: 'manual'` in Fetch to prevent detection of cross-origin redirects.
- Varying the number of strikes necessary for a given domain to be put on the ITP list.

It is important to note that such fixes will **not** address the underlying problem. Specifically, Attack #3 (creating a fingerprint via ITP pins) and Attack #1 (revealing the ITP list) will still be possible due to the large number of application-specific patterns which allow a cross-site attacker to infer the presence of a domain on the ITP list, as discussed in Section 1.2.1.

## 4 ADDITIONAL OBSERVATIONS

While conducting this review, we also noticed behaviors which don't lead to security or privacy vulnerabilities, but which could be interesting and may benefit from a review from the Safari team:

- Bypassing ITP with Public Suffix List domains: A subdomain of a domain on the Public Suffix List[7] (e.g. "foo.githubusercontent.com") is treated as an individual entity for the purposes of ITP. An organization which adds their domain to the PSL can create a wildcard certificate for all subdomains of their PSL domain, and have their cross-site resources loaded from randomly generated subdomains. Since these subdomains are not subject to ITP, requests will contain cookies and the full 'Referer' header, allowing circumventing ITP's tracking protections.
- Non-destructive probing of ITP state: A page can avoid having its cross-site resources taken into account for ITP if the document was opened for less than about 5 seconds. It is helpful for an attacker to be able to check for the presence of a domain on the ITP list without adding a "strike" to the tested domain, for example in order to avoid modifying the ITP fingerprint in Attack #3. We've observed that loads from pages opened for a short amount of time do not add strikes to the ITP list, creating a convenient detection vector.

## 5 ADDENDUM: RESEARCH MOTIVATION

The authors of this report believe strongly in improving the privacy posture of the web and applaud Safari developers' ongoing efforts in this area. At the same time we would like to note that all changes to the web platform that affect its fundamental security properties (such as modifying the behavior of cross-site resource fetches) carry the risk of compromising user privacy and/or security unless special care is taken to understand their impact on the platform. We look forward to collaborating with Apple on future security and privacy improvements to the web.